\documentclass[aps]{article}
\usepackage{geometry}           \usepackage{graphicx}
\usepackage{amssymb}
\usepackage{caption}
\usepackage{subcaption}
\usepackage{amsmath}
\usepackage{fixmath}
\usepackage{MnSymbol}
\usepackage{amsfonts}
\usepackage{array}
\usepackage{multirow}
\DeclareMathAlphabet{\mathpzc}{OT1}{pzc}{m}{it}

\usepackage{caption}
\usepackage{subcaption}

\title{The Slicing Theory of Quantum Measurement: Derivation of Transient Many Worlds Behavior}

\author{Clifford Chafin\\\ \small{Department of Physics, North Carolina State University, Raleigh, NC 27695} \thanks{cechafin@ncsu.edu}}
\begin{document}\maketitle\begin{abstract}
An emergent theory of quantum measurement arises directly by considering the particular subset of many body wavefunctions that can be associated with classical condensed matter and its interaction with delocalized wavefunctions.  This transfers questions of the ``strangeness'' of quantum mechanics from the wavefunction to the macroscopic material itself.    
An effectively many-worlds picture of measurement results for long times and induces a natural arrow of time.  The challenging part is then justifying why our macroscopic world is dominated by such far-from-eigenstate matter.  Condensing cold mesoscopic clusters provide a pathway to a partitioning of a highly correlated many body wavefunction to long lasting islands composed of classical-like bodies widely separated in Fock space.  Low mass rapidly delocalizing matter that recombines with the solids ``slice'' the system into a set of nearby yet very weakly interacting subsystems weighted according to the Born statistics and yields a kind of many worlds picture but with the possibility of revived phase interference on iterative particle desorption, delocalization and readsorption.  A proliferation of low energy photons competes with such a possibility.  Causality problems associated with correlated quantum measurement are resolved and conserved quantities are preserved for the overall many body function despite their failure in each observer's bifurcating ``slice-path.''  The necessity of such a state for a two state logic and reliable discrete state machine suggests that later stages of the universe's evolution will destroy the physical underpinnings required for consciousness and the arrow of time even without heat-death or atomic destruction.  Some exotic possibilities outside the domain of usual quantum measurement are considered such as measurement with delocalized devices and revival of information from past measurements.  
\end{abstract}

\maketitle
\section{Introduction}
The meaning of quantum measurement has been a confounding topic since the early days of quantum mechanics.  The approaches have ranged from very formulaic as in the Copenhagen interpretation to the many worlds view and decoherence \cite{ Everett, Jammer, Schlosshauer, vN}.  The statistics derived from these are typically excellent.  Their accuracy for some systems that have some mix of classical and quantum character is still debated.  Questions about locality and causality regularly arise in the case of correlations \cite{Bell}.  The purpose of this article is to show that a unification of classical and quantum worlds under the same description is easy given the right set of questions and that quantum statistics arise naturally from the dynamical equations of motion (and conservation laws).  Specifically, the sorts of states that lead to observed classical matter arise in a natural way from a primordial delocalized and nonclassical gas due to contraction and the relative cheapness of creating low energy photons.  The photon induced interactions of the induced clusters and massive proliferation of photons, hence increasing dimensionality of the space, will then lead to a kind of ``slicing'' of the space into many classical subspaces in the overall Fock space.  The independence of these are long lasting when their particle numbers are modestly large and slow delocalization is ``resliced'' regularly by the interactions of delocalizing particles with the condensed matter portions of the system.  The small particles that are capable of delocalizing on small time scales are mediators for further partitioning of the space with the probabilities given the square of the amplitude of its wavefunction.\footnote{Here we are referring to the one body wavefunction, $\psi(x)$, that arises from ejection of a localized particle from classical-like matter which will produce a near product function $\Psi_{N}\approx\Psi_{N-1}\psi(x)$ up to symmetrizations.  The framework here will help us extend measurement theory for the collapse of correlated delocalized particles in a causal manner.}  

Any emergent discussion of measurement invariably runs into the need for the many body wavefunction.  This is a high dimensional object and we typically have small particles with delocalization to measure that then interact and produce ``collapse.''  This implies some separability in the net wavefunction.  Any such explanation of quantum measurement must explain the following
\begin{enumerate}
\item The kinds of wavefunctions that correspond to classical matter and their origin
\item The separability of the classical world from the isolated evolving quantum one
\item The statistics of the interaction of the two.
\end{enumerate}
One point often overlooked is that measurements occur at particular times and this is measurable.  A delocalized packet of at atom incident on a surface will give both a location and a time.  Invariably this leads to some vague discussion involving the uncertainty relations, $\Delta x \Delta p\ge\hbar$ and $\Delta E \Delta t\ge\hbar$, however our concern is how the duration of a position measurement relates to the localization in any one slice.  Our goal here is to produce a theory that has no operators or such relations as fundamentals to it.  Rather we seek initial data and an evolution that deterministically arrives at the statistics and evolution we see and, ultimately, gives an explanation for the rather special subsets of wavefunctions that correspond to classical objects and the classical world.  

This article will unfold as follows.  First we discuss a delocalized cooling gas with proliferating photons and how these influence condensing clusters to produce islands of classical behavior for the condensed matter in the many body wavefunction.  These are long lived and promote an arrow of time until the system recontracts and becomes relatively photon poor.  To achieve this we need a description of matter with photon fields of varying number.  Recently it has become possible to subsume the dynamics of QED in a many coordinate and many time classical field theory formalism where the observers perceive a world with \textit{equal times} only \cite{Chafin-QED}.  This formalism and its associated many body conservation laws will be utilized to provide qualitative wavefunction descriptions of measurement as well as quantitative statistics.  Next we discuss how the usual measurement statistics follow for such a system through ``slicing'' over delocalized particle coordinates with such condensed matter states.  A nonlinearity, hidden while using the usual operator formalism, arises in the generation of radiation fields that removes some of the paradoxes in equilibration for purely linear operators on a Hilbert space.  Finally, we use these structures to investigate some paradoxes in quantum mechanics, place some bounds on violation of Born statistics and suggest experiments to reveal such behavior.  

\section{Classical Genesis: A First Look }
The primordial state of the universe is expected to be a gas that cools and condenses into stars and dust.  If the photon number is zero and there are $N$ particles, we expect a single wavefunction $\Psi$ to describe this state.\footnote{We ignore the role of virtual particles to this approximation.}  It is clear, that a general such function is not describable by some mapping to hydrodynamics as a commutative mapping of $\Psi(X)\rightarrow (\rho(x),v(x))$ where the left hand side is governed by the Schr\"{o}dinger equation and the right by Navier-Stokes.  The states on the left are just too large.  Instead of making an argument that the system should settle down to such a state we accept that this may never arise.   It is the author's opinion that classical behavior arises from condensed matter and the proliferation of photons and that it is then induced on gases so we continue our story with nucleation.

Nucleation theory is in still a theoretically very unsatisfactory state and errors in nucleation rates are measure in orders of magnitude.  However, this is fortunately not a complication to the relevant parts of our discussion.  When the atoms of a gas condense into a cluster, a large number of photons are released.  This means that we have now both increased the mean photon number and occupied a large region of Fock space.  The ground state of a cluster of N-particles is nearly spherical (through some polygonal approximation) and rotationally invariant.  This seems initially paradoxical.  No discrete crystal has rotational invariance.  The resolution follows from the fact that these are $3N$ dimensional wavefunctions.  The translation is given by three of these and the rotational freedom by two more.  Rotation always requires radial excitation, as we see from the case of the Hydrogen atom.  In the case of a large cluster, this radial excitation is a centrifugal distortion.  The rotationally invariant ground state has no well defined atom location, even if the structure is crystalline in that we cannot find peaks at locations $r_{i}$ so that $\Psi\sim\prod_{S}(x_{i}-r_{j})$.  The states where such arises, as in the physical states we observe, must then be manifested by the cluster being in a mixture of high rotational eigenstates (even if having net angular momentum zero).  

A surprising complication is that any classical body is in such a mixture of states so, even at ``$T=0$'' it is far from its own ground state.  The kinds of condensed matter we encounter have well defined shape, orientation, etc.  They define a ``classicality'' that is very specific, three dimensional and Newtonian, and far-from-eigenstates.  A solid can be specifically described and phonons given as excitations of the localized cores along particular many body diagonals and are eigenstate-like despite the ultimately transient nature of the classicality on which their description depends \cite{Chafin-auto}.  We now are compelled to ask how such apparently omnipresent states can arise.

Consider a pair of irregularly shaped bodies, A and B, that are spatially separated, but suffering delocalization about their centers of mass, and are bathed in a sea of photons.  Let these be in their ground states initially.  A photon that travels from far away and casts a shadow from body B onto A gets absorbed and produces a localized excitation on them.  In the case of absorption by A the surface builds up a history through local heating or chemical changes.  After many such photon events the body A has a record of the shape of body B in this shadow.  Of course, some fraction of the amplitude of each photon gets absorbed by B or flies past without interaction.  If the bodies A and B had localized atomic constituents, then their boundaries would be well defined and the shadows sharp.  Since this is not the case we have to ask what happens.  We can consider each to be a superposition of states that are in various angular orientations.  This is reasonable since the centrifugal forces of these many angular states are small and make little deformation of the bodies.  Each such case produces shadows that are well defined so we have a macroscopic superposition of all the configurations with well defined orientations and atomic locations.  The crucial part is how this then evolves.  

Given a superposition of nearly overlapping macroscopic bodies in a space with no photons the energy change is huge.  Atoms cannot sit on top of each other without inducing large repulsive forces from their electronic structure.  However, for a system with a huge variation in the photon number states, such slight changes can easily have different photon numbers so be, ostensibly, at the same location but in different photon number spaces.  This allows an apparent overlap with no energy cost.  Specific details of this rely on an initial value (rather than operator based) description of low energy QED described in \cite{Chafin-QED} and summarized below.  Since the delocalization rate of large $N$ objects is very small, such states can then evolve for long periods of time with essentially no interaction between them.  Ultimately, we are such objects.  Our very consciousness and memory depends on our being reliable discrete state machines.  Once the expanding and cooling universe is so partitioned we have a set of ``many worlds'' that are sufficiently separated in Fock space to be insulated from each other.  Of course, this is not expected to persist.  In a gravitational contraction or long term stagnation, these worlds will come back together and the ``information'' made up by these separated worlds will be lost.  This is an appealing way for the arrow of time to arise naturally despite the time reversal symmetry of the equations of motion.  To be fair, this is a very vague and qualitative discussion.  Now let us try for a more specific, but less general case in an attempt to justify this partitioning of the many body wavefunction.  

\section{Classical Genesis: Cluster Collisions and Photons }

Here we give a justification for the ``sparse worlds'' state that we claim is a set of many-body wavefunctions that correspond to classical condensed matter objects (plus gas and a few delocalized particles).  By this we mean that the solid and liquid objects have well defined boundaries, shapes and orientations as 3D objects but encoded in the N-body space of atoms where these atoms have well defined locations to within some localization distance determined by the electronic bonds between them.  Of course, such a state is not an eigenstate.  Each body will tend to delocalize both radially and in location.  Such a state is an unfathomably complicated mix of eigenstates of the true system yet it makes some sense to think of the excitations of the bodies in terms of collective phonon modes as eigenstates in such clumps of matter.  

Matter begins in the universe as a gas that collapses into stars and explodes to create the clusters that condense into dust that eventually coalesces into planets and other rocky objects.  The gas undoubtably begins as delocalized and ``correlated'' in the sense that the particles have no well defined 3D locations so the many body $\Psi$ cannot be represented as some symmetrized N-fold product.  The implications of this are rarely considered.  How does classical hydro arise in such a system and lead to stars of well defined location much less the larger scale density structures we observe?  Is this classical localization a result of some product of our consciousness in creating a ``measurement.''  This is pretty unpalatable to most scientists.  The alternative is that such condensing occurs but the resulting stars have no well defined location, particle number, boundary and orientation relative to one another.  Such a universe is a truly many body object and how it would ``look'' to an observer injected into it is not clear.  Later we will see that the consciousness required for observation may be incompatible with such a universe.  

The resolution we suggest is that this is the true state of the early universe and it is the presence of condensed matter that ``slices'' the space into a well defined collection of stars of well defined locations and velocities.  The collapse picture implies that only one such state is selected and exists.  In this picture,  the the coordinates of the observer contain copies of the ``observer $\otimes$ system'' that cease to be the same for all values of the system coordinates.  This divides the wavefunction of the many body space into a collection of independently evolving states of well defined 3D structure with long lasting independence and duration.  We can then think of quantum measurement as the ``auto-fibration'' of the macroscopic world over the coordinates of the measured particle.  

Consider a classical-like block of matter floating in space.  
A superposition of a star at two locations shining on such a block creates a superposition of the block in the star's coordinates.  If we view the block as a measurement device that is recording observations in the changes in its surface under the influence of photons from the star, then it
 ``observes'' its own history to have the star at one continuously connected path of locations.  It now has a double life as two blocks with different histories even though the number of coordinates has not changed.  It's classicality has been compromised (albeit in a very minimal way) by the influence of the delocalized star even though the star and the block are widely separated and the net mass and energy transferred by the photons is typically miniscule.  The ``measurement device'' has not forced a change in the larger system.  Rather, the larger system has induced a change in the measurement device so it now follow separate paths in the many body space.  This is possible, in part, due to the massive size of the many body space and its capacity to hold many classical world alternatives as distinct for long times.  Note that the size of the block compared to the superimposed object is irrelevant in producing this effect.  

The problem them amounts to the creation of such a set of classical-like bodies distributed in a set of sparse worlds embedded in the many body space.  As a prototype world consider a collection of dust of different sizes, shapes, orientations, internal excitation, positions and velocities.  These begin as a highly correlated system that has no classical meaning despite having formed solid matter.  Let us start with an idealized simple system to discuss the mechanism.  Consider two solid balls of radius $r$ but nonspecific location and velocity in many body space described by a cube of length $L$.  Ignoring internal degrees of freedom, we can consider the system to be a 6D wavefunction in an $L^{6}$ cube with excluded volume given by the 2 body cylindrical projection of the interior of the sphere.  At higher energies the wavefunction will tend to have oscillations much smaller than the radius $\lambda\ll r$.  The state of the system in terms of eigenstates is assumed to be of a broad energy distribution $\Delta E\gtrsim~\!\! <E>$ and have random phases or have evolved for a long but random length of time.  Such a condition is necessary to have fluctuations in the many body current $\mathcal{J}$.  The energy density and fluctuations then tend to uniformly fill the box and we have a soup of high frequency and highly varied oscillations bound by the excluded volume.  

So far we have said nothing about photons.  Let us assume there are none to start with.  As currents induced by the wavefunction produce flux on the boundaries of the excluded volume.  Classically this corresponds to the collision of two spheres with velocities given by the two velocities $$v_{1},v_{2}=\mathcal{J}/\mathcal{P}$$ given by the 6D current $\mathcal{J}$ and density $\mathcal{P}$ at the coordinate $X=(x_{1},x_{2})$.  Depending on the angle and relative speed of the collision, a certain number of photons care created in the event.  Photons are exceedingly inexpensive at low energies.  This has led to the infrared divergence problem in QED where an unbounded number of low energy photons get created.  Our finite box regularizes this to some degree but for short enough collision times no such problem arises since they cannot traverse the box during their creation.  

A small change in the location of the collision creates a different number and set of photons.  Thus one location can generate a large occupancy in the tower of spaces $\Psi_{bb}, \Psi_{bbA}, \Psi_{bbAA}, \Psi_{bbAAA}, \Psi_{bbAAAA}\ldots$ where $b$ indicates the coordinates of each ball and $A$ are the photon coordinate labels.  In a short time, the current flux at that location can be very different and generate a very different occupancy the the ball-photon wavefunction tower (Fock space).  Once each small current fluctuation is completed, the higher photons spaces have acquired an occupancy of localized spatial position in the $b$-coordinates (defined by the length of time of the local fluctuation in current) and a broad number of photon waves moving away from it in the $A$-coordinates.  The long time limit we argue is of a sum of such states distributed among the tower with almost all the amplitude having left the $\Psi_{bb}$ state.  These can now evolve with no quantum interference of other states (since all $b$ and $A$ coordinates would have to match up in one of the towers for this to happen).  By ``long time'' we mean long enough for the currents in the $\Psi_{bb}$ state to have had time to have all reached the excluded volume surface and hence pushed amplitude up the photon tower, $\tau\gtrsim L/\text{Min}(v_{1},v_{2})$, but not so long as to cause delocalization of the amplitude in each n-photon space so these begin to interact and interfere.  

The actual process ``in vivo'' of the universe is of course more organic and occurs while the dust is forming.  It must create the orientation of the dust as well as select these subslices to have well defined atom number in each.  It seems that the cheap and plentiful photon along with dust formation is what drives the formation of these ``classical worlds'' as isolated long lasting packets in the many body space.  Quantum mechanics then arises for each of these universes by the action of condensed matter as discrete state machines.  Clearly this process cannot persist forever.  The universes will delocalize, meet, possibly gravitationally collapse and get driven to a density where the full correlated structure of the universe matters.  


\section{Measurement}
Part of the formalism of quantum mechanics has been to use Hilbert space and eigenfunctions of operators to give measurement results.  These Hamiltonians are often effective Hamiltonians of subspaces created by the kinds of localized ``classical'' states described above.  This introduces a kind of metastable feature to the evolution that is connected with the duration of the classical nature of the external world.  One has to wonder what the role of the eigenstates are in arriving at measurements, specifically how one collection of matter indicates one particular operator and spectrum.  In the case of position measurements, we see from above that the system has partitioned itself so that measurement of particle location is inherited by the special independently evolving nature of the classical states.  In this case we say the system has been ``sliced'' in a manner that gives it its classical character but not into a subset of eigenstates of the net or any obvious subset of the Hamiltonian.  We assert that momentum, energy and other measurements are universally inferred from position data e.g. a local color change in a material or spatial measurements at different times.  It has already been long debated how general a measurement can be made from an arbitrary linear self adjoint operator (LCAO) and it is this author's opinion that position and time measurements are the fundamental sort that arise and all others are derivative.

Note that our ``measurement'' process has nothing to do with consciousness of an observer but of a specific property of condensed matter in a photon rich environment.  In fact, photon production at low energies is so cheap that it is hard to conceive of a measurement that didn't produce copious numbers of them.  Let us now consider temporal effects and measurements.  It is inevitable that temporal effects arise.  Wavepackets can be delocalized and measurement devices can move.  This makes it clear that the measurement operator $\hat{x}$ is going to have some insufficiencies.  Furthermore, measurement devices have finite spatial extent.  Screens are essentially 2D so they are typically only picking up a tiny fraction of a wavefunction's motion at any time.  

\begin{figure}
  \begin{centering}
 \includegraphics[width=2in]{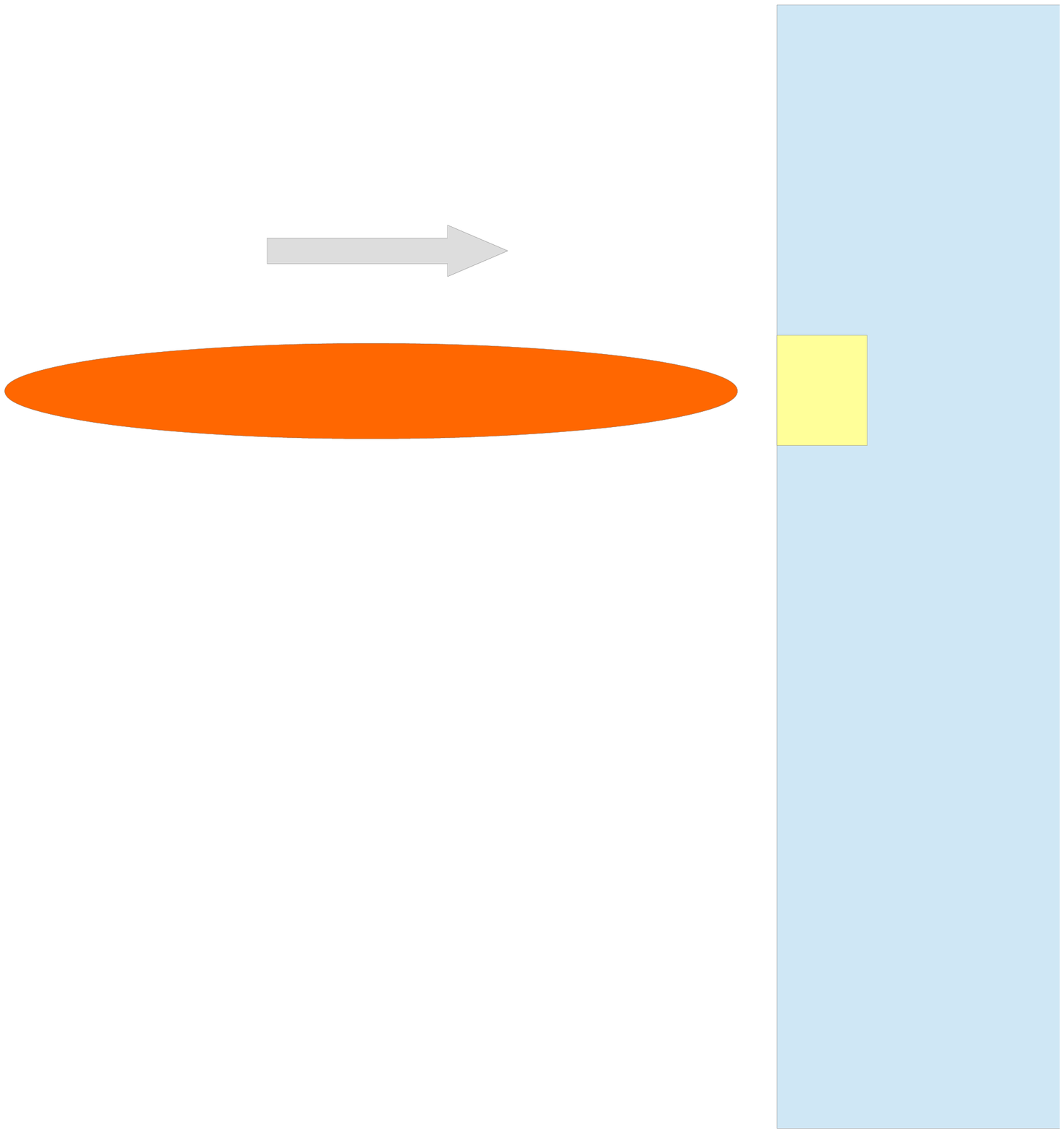}%
 \caption{\label{long} A long narrow packet illustrates the measurement of event time at a particular location and how these can lead to a persistent slicing of the space (up to the delocalization time of the device) in an infinite space.}
   \includegraphics[width=2in]{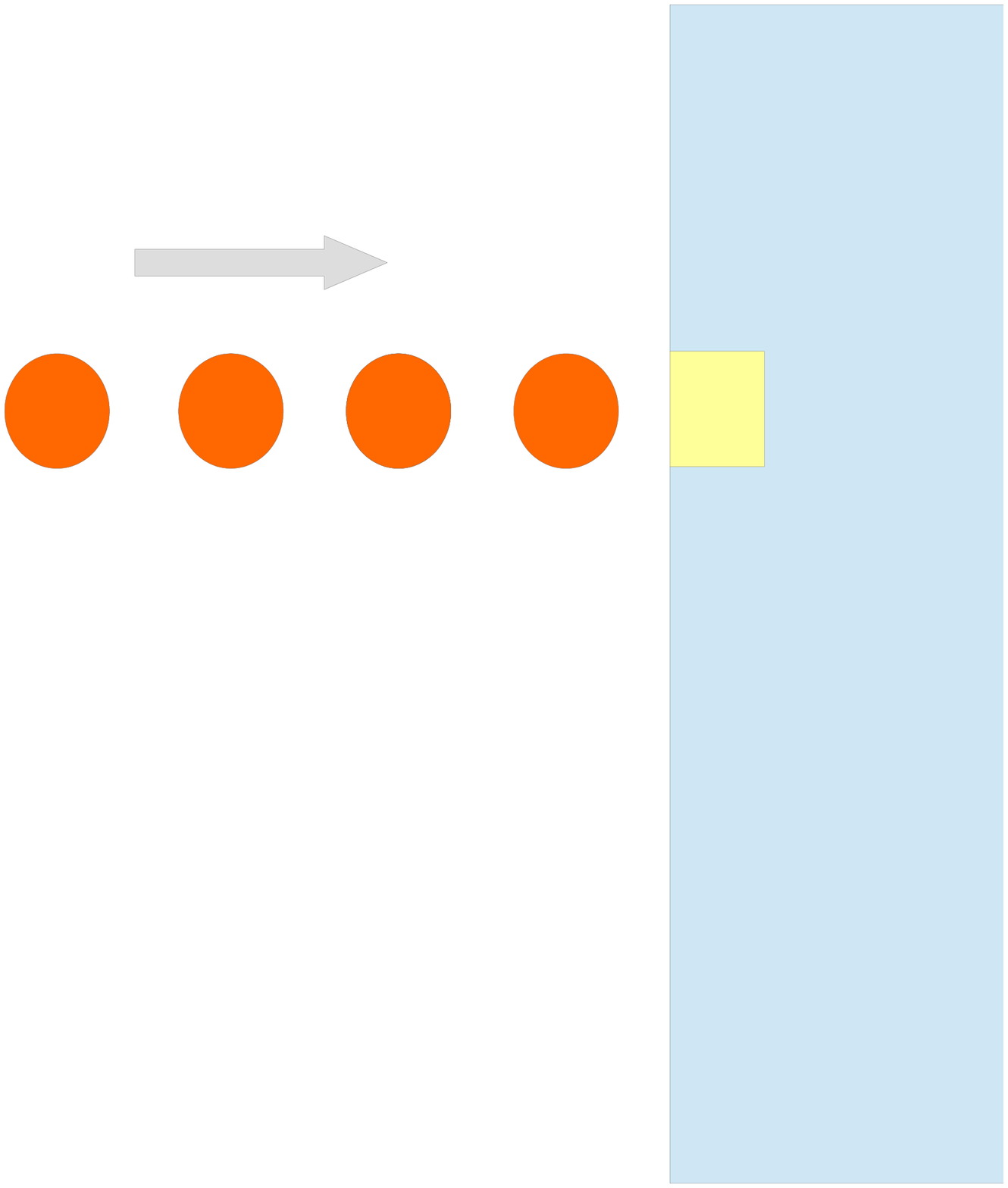}%
 \caption{\label{pulse} An idealized sequence of packets of a \textit{single} incident particle.  }
 \end{centering}
 \end{figure}

To illustrate these points consider a narrow single particle packet incident on a screen with a couple of adsorption sites as in fig.\ \ref{long}.  We can simplify this by breaking it up into a set of disjoint regions of support as in fig.\ \ref{pulse}.  The duration of an adsorption event is not related to the length of a packet but the radiation time for the electronic decay that produces binding.  For simplicity let the binding action be mediated by the release of a single photon of energy $E$ so the radiative process has a time scale $\tau\sim\hbar/\Delta E$.  Let the parcels be roughly monochromatic so they have a well defined velocity velocity $v=j/\rho$ and the parcel widths $w\approx v \tau$.  A parcel separation of $nw$ lets the adsorption events be well separated.  

When a subparcel reaches the site at $x_{0}$ it adsorbs and creates a photon so that some amplitude flows from $\psi(x)\Psi_{N}$, the photon free wavefunction of the system, to $\Psi_{N+1,A}$, the single photon and N+1 particle wavefunction with a radiation field flowing away from it.  The operator formalism obscures some features of this problem so we invoke an equivalent formalization of low energy QED by using a many time approach where one body equations of motion hold for each time coordinate in the many body tower \cite{Chafin-QED}:   
\begin{gather}\label{tower}
\vdots\\\nonumber
\Psi_{N,AAA}\\\nonumber
\Psi_{N,AA}\\\nonumber
\Psi_{N,A}\\\nonumber
\Psi_{N}\nonumber
\end{gather}
We call such a theory  ``deterministic wave mechanics'' (DWM) in contrast with the formal operator and path integral formulation of the theory.  
A basis of states in each photon number space is given by 
$\Psi_{N}^{(m)}\mathcal{A}_{m}$ where $\mathcal{A}_{m}$ is a stationary state in the space spanned by $A^{i_{1}}\otimes A^{i_{2}}\otimes\ldots\otimes A^{i_{m}}$ of complex 3-vectors fields for photons.\footnote{Coulomb gauge is assumed for every coordinate label so that the $\Psi_{N,1}^{\mu=0}$, $\Psi_{N,2}^{\nu,\mu=0}$, etc.\ components are fixed by constraint.}
The net norm and energy are conserved in such approach when they are defined as
\begin{align}
\hat{\mathcal{N}}(\Psi_{N,n})&=\int dx^{i_{1}}_{s}\ldots dx^{i_{N}}_{s} \bar\Psi_{N}\Psi_{N}\\ &+ \frac{1}{4\mu_{0}}\int dx^{i_{1}}_{s}\ldots dx^{i_{N}}_{s} \int dx^{i_{1}}_{A}\ldots dx^{i_{n}}_{A}\sum_{k=1}^{n} \left( \bar\Psi^{i_{1}\ldots i_{n}}\partial_{t_{A}^{i_{k}}}\Psi_{{i_{1}\ldots i_{n}}} -\partial_{t_{A}^{i_{k}}}\bar\Psi^{{i_{1}\ldots i_{n}}}~\Psi_{{i_{1}\ldots i_{n}}}\right)\\
&=\int dx^{i_{1}}_{s}\ldots dx^{i_{N}}_{s} \bar\Psi_{N}\Psi_{N}\\
&+
\frac{1}{4\mu_{0}}\int dx^{i_{1}}_{s}\ldots dx^{i_{N}}_{s} \int dx^{i_{1}}_{A}\ldots dx^{i_{n}}_{A}\sum_{k=1}^{n} \left( \bar\Psi^{i_{1}\ldots i_{n}} \hat{\mathcal{N}}^{A}_{k}\Psi_{{i_{1}\ldots i_{n}}}\right)
\end{align}
\begin{align}
E_{N,k}=\bar\Psi_{N,k}\left(\sum_{i=1}^{N}\hat{E}_{s_{i}}\hat{\mathcal{N}}_{1\ldots \hat i \ldots N}\hat{ \mathcal{N}}^{A}_{1\ldots k}+\sum_{j=1}^{k}\hat{E}_{A_{j}}\hat {\mathcal{N}}_{1\ldots N}\hat {\mathcal{N}}^{A}_{1 \ldots \hat j \ldots k}\right)\Psi_{N,k}
\end{align}
and we evaluate on the equal time slices $t\doteq t_{net}=t^{i_{1}}_{s}=t^{i_{2}}_{s}=\ldots=t^{i_{1}}_{A}=t^{i_{2}}_{A}=\ldots$.  The operators $\hat{\mathcal{N}}_{s}$ and $\hat{\mathcal{N}}_{A}$ are the one body norm operators for massive and photon fields respectively.  The operators $\hat{{E}}_{s}$ and $\hat{{E}}_{A}$ are similarly the one body energy operators.  The many body versions are simply concatenations of these where the ``hatted'' indices are excluded.  The definition of $\bar\Psi$ for Dirac fields is to apply $\gamma^{0}$'s to all the spinor indices of $\Psi$ (which have been suppressed here). Here we are interested in atomic center-of-mass wavefunctions.  For these we simply require the transpose conjugate.   

\begin{figure}
  \begin{centering}
 \includegraphics[width=2in]{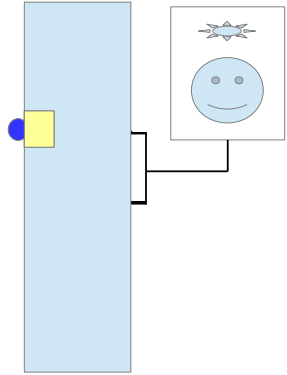}%
 \caption{\label{observer} A measurement device with a coupled observer or programmable device to respond to observations. }
 \end{centering}
 \end{figure}

\begin{figure}
  \begin{centering}
 \includegraphics[width=2in]{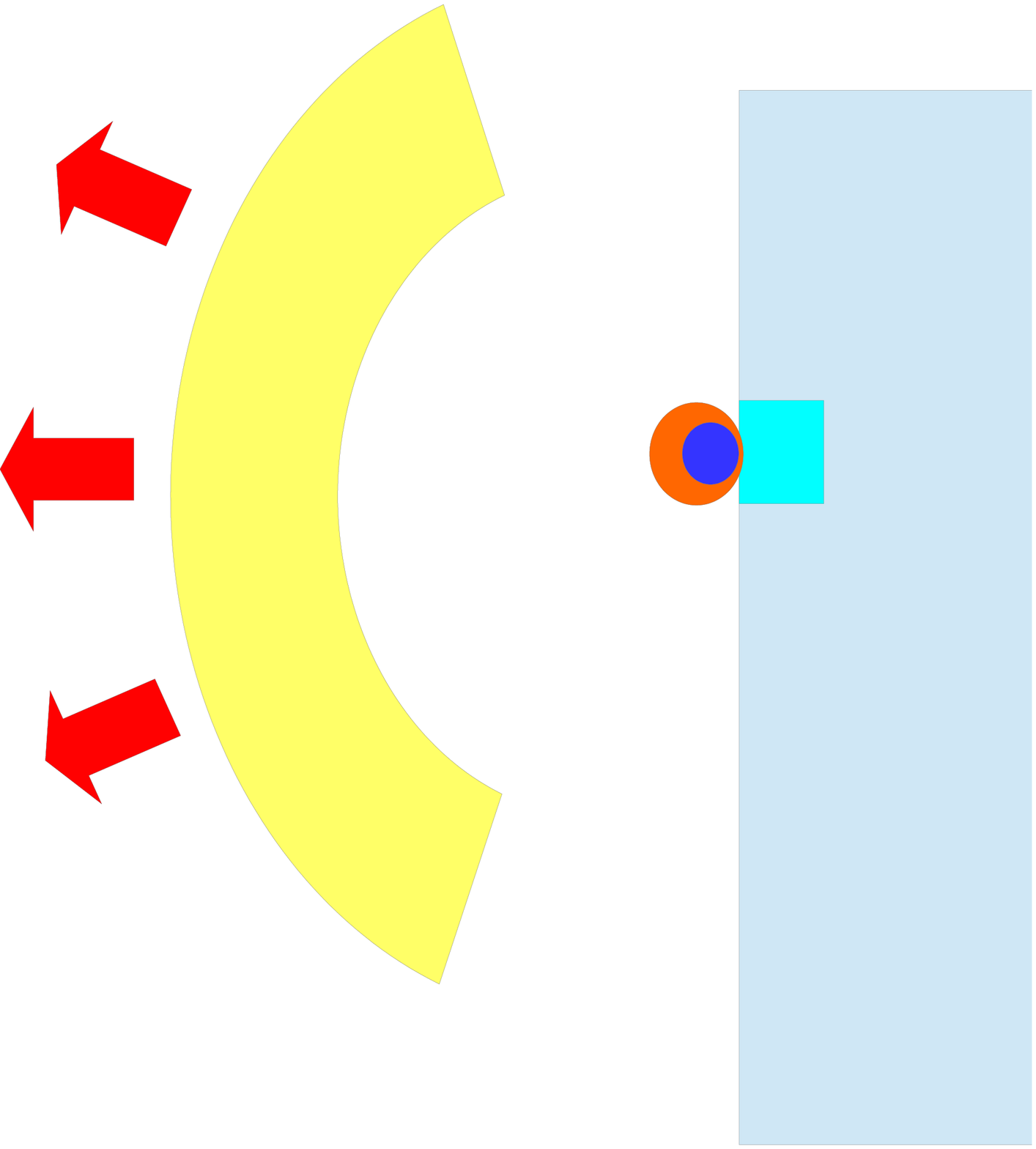}%
 \caption{\label{radiation} The absorption of a particle at a site is correlated with radiation field moving away from the selected location. }
   \includegraphics[width=2in]{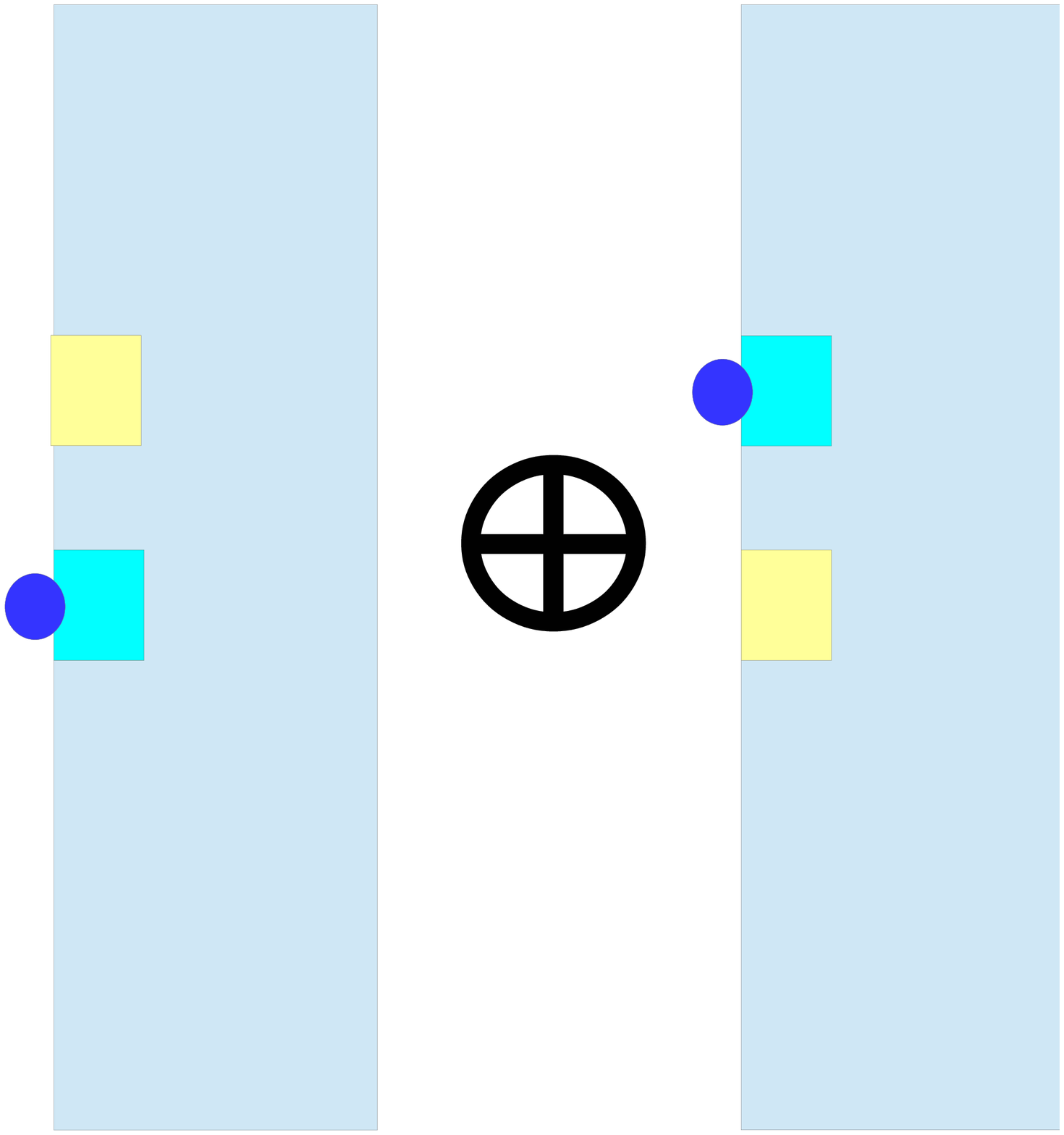}%
 \caption{\label{product} The two possible configurations of a broad packet measurement (with suppressed radiative fields) exist as a kind of direct sum indexed by the coordinate label of the original incident particle.}
 \end{centering}
 \end{figure}

Using this picture we can derive the long time states of the system.  The radiative decay occurs at frequency $\omega$ with an envelope of duration $\tau$ as in fig.\ \ref{radiation}.  The atom binds a location $x_{0}$ with a mean width of $d$ so that it may be represented by a peaked function $\delta_{d}(x-x_{0})$ akin to a delta function of finite width $d$.  Assume the first peak arrives as time $t=0$ and that there are only two equal pulses that contain all the amplitude of $\psi$.  Initial data at $t\lesssim 0$ is 
\begin{align}
\Psi_{N+1}&=\Psi_{N}\psi(x,0)=\frac{1}{\sqrt{2}}\Psi_{N}(\delta_{w}(x-x_{0})+\delta_{w}(x-x_{0}-wn))\\
\Psi_{N+1,A}&=0\\\vdots
\end{align} 
The final wavefunction for $t>t'=2\tau+n\tau$ is
\begin{align}
\Psi_{N+1}&=0\\
\Psi_{N+1,A}&\approx\frac{1}{\sqrt{2}}\Psi_{N}\delta_{d}(x-x_{0})\left(\frac{e^{i(kr-\omega t)}}{r}h(r-ct) +\frac{e^{i(kr-\omega (t-t'))}}{r}h(r-c(t-t'))\right)e^{i\phi(t)}\hat{\epsilon}_{\kappa}\\
\Psi_{N,AA}&=0\\
\Psi_{N,AAA}&=0\\
\vdots
\end{align} 
We have implicity assumed the block is essentially transparent and the radiation flies unobstructed into infinite space.  (The orientation of the radiation field $\hat\epsilon_{\kappa}$ is determined by the direction of the dipole produced by the radiation.  This may be a superposition of such solutions and a function of the local geometry of the solid.  For now we neglect its details.)  The meaning of this solution is that the wavefuction support has exactly partitioned into two parts.  The ``reality'' of a classical field can have some surprising subtleties \cite{Chafin-unif}.\footnote{We can consider this as the ``Schr\"{o}dinger'' and ``first quantized'' analog to usual QFT formalism in terms of field operators.}  In this case the support and its values there contain all the meaning there is to the system.  We see that we have two bound states that occurred at times $t=0$ and $t=t'$.  The packet is flying away from the location $X\approx x_{0}\otimes X(0)$ at $c$ in the $x$ direction when viewed in the equal times coordinate $t$.  The motion in the material coordinates is essentially static unless some other dynamics were present to start with.  If we consider the block to contain a discrete state machine as in fig.\ \ref{observer} that has internal dynamics that makes a record of when the event occurs, then each one exists in a kind of parallel universe with a record of a different time.  Unless these photon coordinate portions of the packet are reflected or forced to interfere, this situation continues in perpetuity and each evolves according to their own record of their particular past.  Should they generate their own delocalized particles and repeat this experiment they will find the Born-like $\psi^{*}\psi$ probabilities for when the measurement occurs.  This is a direct consequence of the above norm conservation law.  Ultimately the delocalization can only go on so long before the ``classicality'' of the system fails.  The consequences of this we will soon consider.  

\begin{figure}
  \begin{centering} 
\includegraphics[width=2in]{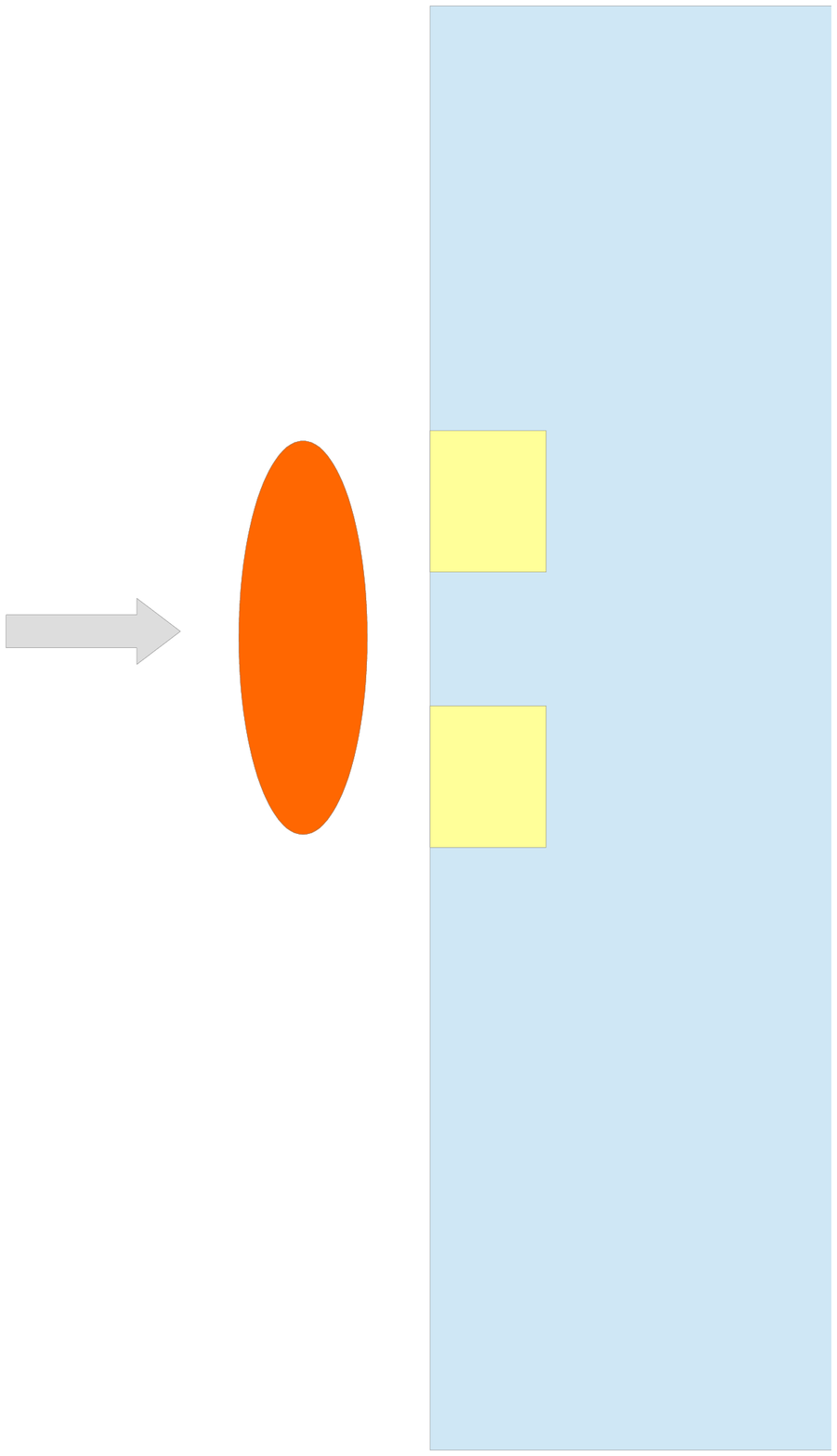}
\caption{\label{broad} A narrow one-particle packet incident on a detector surface.}
\includegraphics[width=2in]{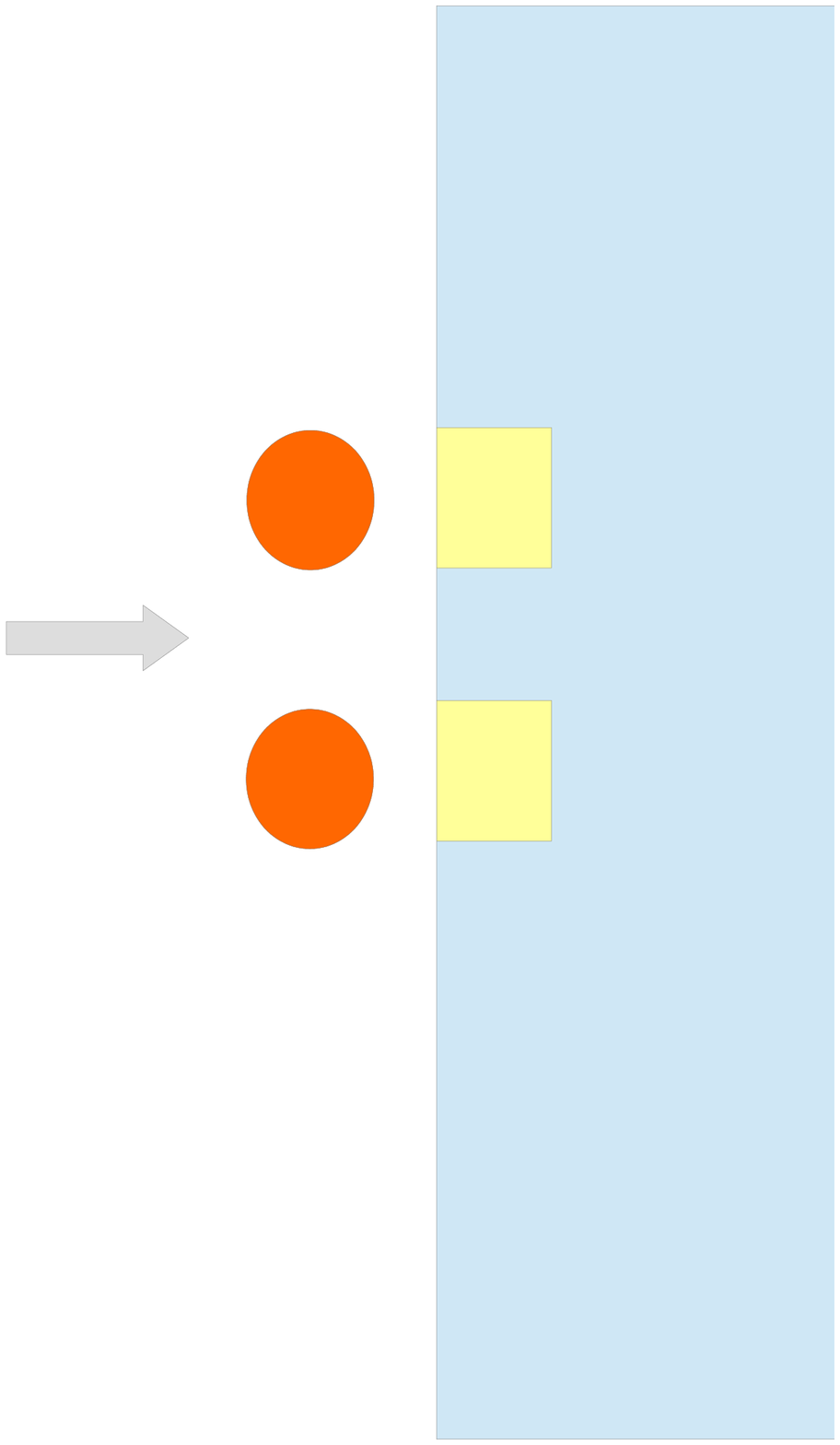}
\caption{\label{side} An idealization of the narrow one-particle packet into localized subparcels.}
\end{centering}
 \end{figure}

Let us now consider a broad packet that intercepts the screen at the same time as in fig.\ \ref{broad}.  Analogously to above, let us consider this to be broken into two parts with the width of the measurement centers and less than $w=v \tau$ as in fig.\ \ref{side}.  Here a similar analysis yields a resulting pair of packets radiating outwards from the two centers at the same time.  Our system now seems to be split into two spatially distinct parts as indicated by the outer product in fig.\ \ref{product} where the radiative field shells have been suppressed.  These shells are no longer disjoint but contain a finite volume fraction of overlap.  For farther apart centers this is of order $w/R(t)$ where $R(t)=ct$.  To the extent this overlap remains negligible, these solutions remain disjoint and evolve as separate worlds.  

This is a good point to pause and reflect on what overlap of these systems means for evolution.  The emphasis on linear operators and Hamiltonians leads one to believe that any superimposed world is equivalent to each world evolving separately.  As such, when one decomposition evolves it is hard to see how anything interesting can really happen.  However, there is a hidden nonlinearity in our problem.  The classical radiation reaction problem holds a nonlinearly due to current acceleration which is best thought of in terms of finite sizes of radiators and crossing times \cite{Rohrlich}.  Our radiation fields can be thought of in a similar fashion with a small unknown structure involving many hidden internal coordinates.  The ``radiation reaction'' now must transfer both four momentum \textit{and} particle norm at the interacting two-body diagonals that connect the states in the Fock space tower.  The implications of this is that overlapping of states in the Fock space do not simply superimpose so there are no true eigenstates when photon interactions are included.  This is to be expected.  If we superimpose the eigenstates $\psi_{2p}$ and $\psi_{1s}$ of the Hydrogen atom then it is the presence of the current that drives amplitude from $\psi_{H}$ to $\psi_{H,A}$.  In the low energy limit the Hydrogenic states are stationary but the overlap drives the transition to higher photon levels.  This is an intrinsic nonlinearity that is obscured by the formal operator description of quantum field theory.  It is unclear if this is adequately accounted for in quantum field theory through its operator calculus.  

\section{Slice Memory and Revival of Measurement History}
One of the unpleasant features of the many worlds interpretation is that the size of the universe seems to grow.  In this and all ``interpretations'' of quantum mechanics, the role of the measurement device and how and when it acts lacks specificity.  The action of the ``observable'' associated with each such device is not clearly determined by the microstructure of the device.  The DWM theory here addresses each of these and lets us ask some new questions that may take us outside the bounds of traditional quantum theoretical problems.  One of the obvious questions is to what extent is the measurement a complete destructive event (at least from the perspective of the observers).  Can we somehow undo measurement and recover some of the delocalization and phase information from before?  Now that we can nanoengineer systems and create extremely cold ones, highly decoupled from the external world, other quantum domains can be probed.  A molecular two-slit experiment was recently realized \cite{Kreidi}.  
In the measurement direction what happens when a measurement device itself has a mass comparable to the delocalized system it measures?  Is there a measurable ``back reaction'' to the measurement event?  If a measurement device is partially delocalized itself how does this affect the measurement once we then slice the measurement device so it is back in the fully classical domain of our experience?  

\subsection{Wavefunction Revival: Inverse Measurement}
On the topic of slicing of the space into independently evolving subspaces we have introduced the restriction on the form of macroscopic matter that gives a classical limit for dynamics.  This was far more restrictive than the rather naive Ehrenfest-limit defined by large mass and moving packets \cite{Sakurai}.  The continuing lack of overlap given by large mass induced slow spreading and the rapid motion of light speed packets in the A-coordinate directions into an empty space help preserve this ``many-worlds'' picture for long times.  Constraints on the space that photons can move about in leads to greater overlap possibilities and opportunities for such slices to interact through radiation absorption and production however, since low energy photons are so prolific this kind of interference may be difficult to engineer in practice.  Nevertheless, we should investigate the possible bounds on slice independence.  

Consider the example system given in fig.\ \ref{side}.  Generally, there are going to be internal motions and radiation fields that exist in any such large body.  Let the incident atom be distinct from those of the device so that it is unconstrained by symmetry and the binding to the surface can be much less than that of the device particles to each other.  We can imagine a situation where we heat the block and the atom ejects and delocalizes then in pulled back the the surface by an external field such that this process is iterated.  The CM of the device gradually delocalizes (at a much increased rate) from this process.  If this system is closed then the photon number will gradually increase as the battery driving the process loses energy.  This tells us that the system is undergoing important changes and so reejecting the particles may not create a system that interferes with previous slice histories.  On the other hand, if the system is in a finite volume, the radiation fields can all be contained in this finite space so that past slices eventually can interfere if the photon number does not grow much faster than the number of iterations.  

It is simpler to consider the case of a photon that is absorbed at a pair of sites and then ejected as in the process $\Psi_{N,1}\rightarrow\Psi_{N,0}\rightarrow\Psi_{N,1}$.  The release times for the two slices may vary over a large range but, if we restrict ourselves to looking at the fraction of amplitude that occur at the same time (e.g. by use of a beam chopper on the input and ejected flux), then the phases of the resulting two components of the single photon may be compared.  After absorption, the system is a photon free wavefunction consisting of a superposition of two different internally excited states that evolves according to the net mass-energy in it.  The relative phase of each space is fixed by the phase difference of the original photon at the time it was absorbed by the two sites $\Delta\phi=\phi(x_{1},t=0)-\phi(x_{2},t=0)$.  Restricting our measurements to the case where the frequency of the emitted photons are the same, this phase difference should be preserved in the $T=0$ limit.  Thermal fluctuations in phase between the two points will produce a shift in this value.  This procedure gives a measure of the regional phase fluctuations and isolation of the system.  

\subsection{Measurement Back Reaction}
The subject of back reaction has been around for some time \cite{Holland}.  If one believes in a collapse picture then one can readily see that center of mass motion is not conserved in a position measurement.  This means either it is truly not conserved or there is an unspecified back reaction on the system.  In DWM we see that conservation laws only hold for the totality of slices not for individual ``observer-paths.''  Therefore no back reaction is expected.  We can utilize a pair of ultracold traps to give a specific test of this.  Given a delocalized large mass molecule in a pair of widely separated traps we can send an atom through two paths to make contact with each of these.  If a collapse produces a net conservation of all the usual conserved quantities then the center of mass shift will be proportional to the separation of the traps so can be made as large as desired and easily detected by florescent behavior of the molecule.  

\subsection{Nested and Fuzzy Measurement}
The meaning of superposition of macroscopic objects has been debated at least as long as the famous Schr\"{o}dinger's cat paradox \cite{Schrodinger, Carpenter}.  By our judicious selection of initial data we see that this is resolvable.  The overlap of such states is explained by the proper consideration of correlations of photon fields in partitioning the system under such a slicing event as above.  The nature of macroscopic superposition does however beg some interesting questions when the measuring device is also delocalized.  For example, if the incident $\psi$ has positive and negative regions that are shared equally over the same site due to delocalization then the net norm of $\psi$ at that site may be zero.  Does this mean there is no probability of adsorption at the site and the amplitude there is reflected?  Furthermore, we can ask if the order of a meta-observer's action on the system in measuring the measurement device before it acts on the $\psi$ or after makes any difference in the resulting statistics.  These two scenarios can be classified as ``fuzzy measurements'' and ``nested measurements.''  

Firstly, consider a ``device'' that is a pair of separated, localized and slowly spreading heavy atoms or molecules in a trap.  This allows for the possibility of the larger bodies capturing a small atom then moving the bound bodies around before ejecting the light atom from them.  If the atoms are initially well localized and remain so for the duration of the experiment then the resulting phases on revival will be determined by the amplitude emission time and rate from each source atom.  Note that this situation depends on the particles and what is moving them.  If they are isolated like a gas then this is certainly true.  If, however, the particles are being localized and moved by macroscopic classical matter or radiation that then is absorbed by it then the interactions with the external world may produce a slicing of the system.  There may be no ``meta-observer'' or other unsliced mechanism to eject the light atoms and produce a spreading in its coordinate direction that causes the system to be seen as a wavefunction with some stored phase history and an external world.  We can apply a radiation field to eject the light atom but have no way to know that our counterparts in the other slices have chosen to do the same.  

\begin{figure}
  \begin{centering}
 \includegraphics[width=2in]{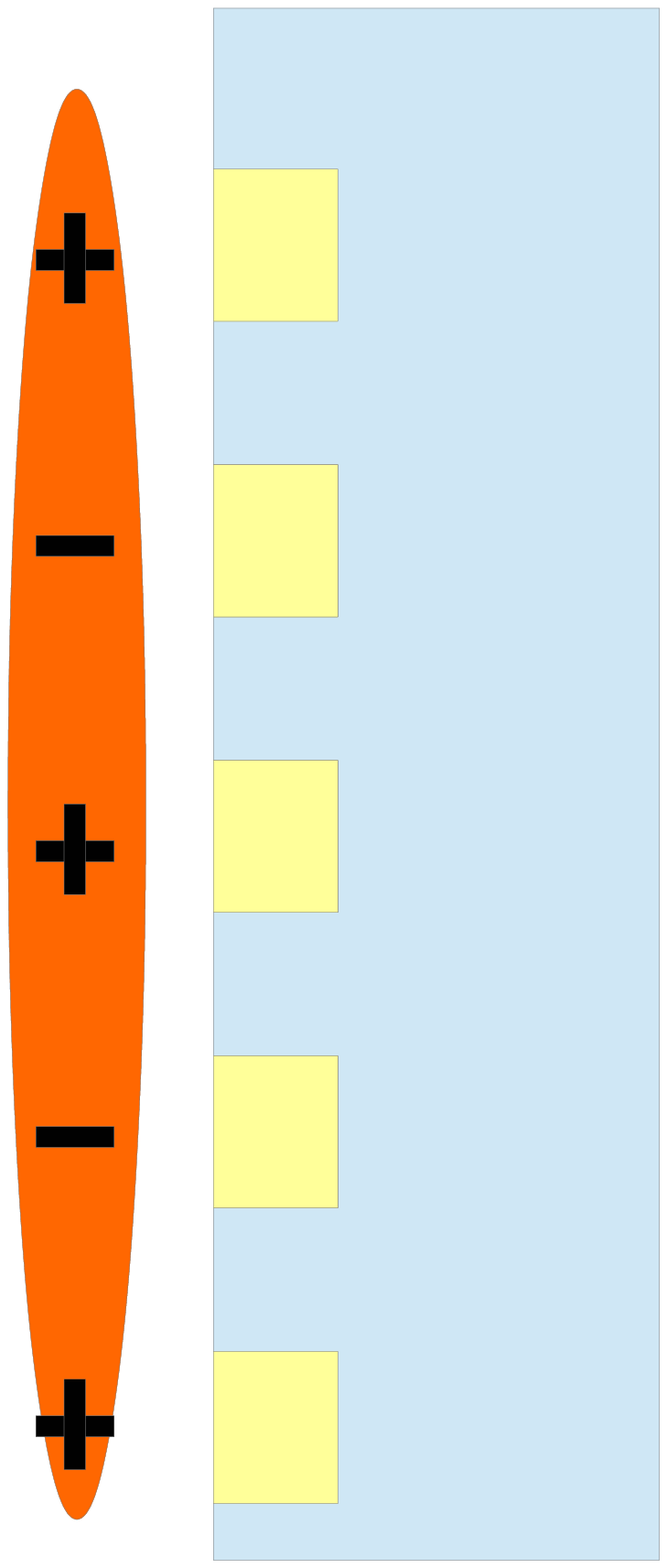}%
 \caption{\label{verybroad}  A broad narrow packet incident on a screen.  There is a relatively slow phase oscillation component parallel to the surface that matches the possible adsorption sites.}
  \includegraphics[width=2in]{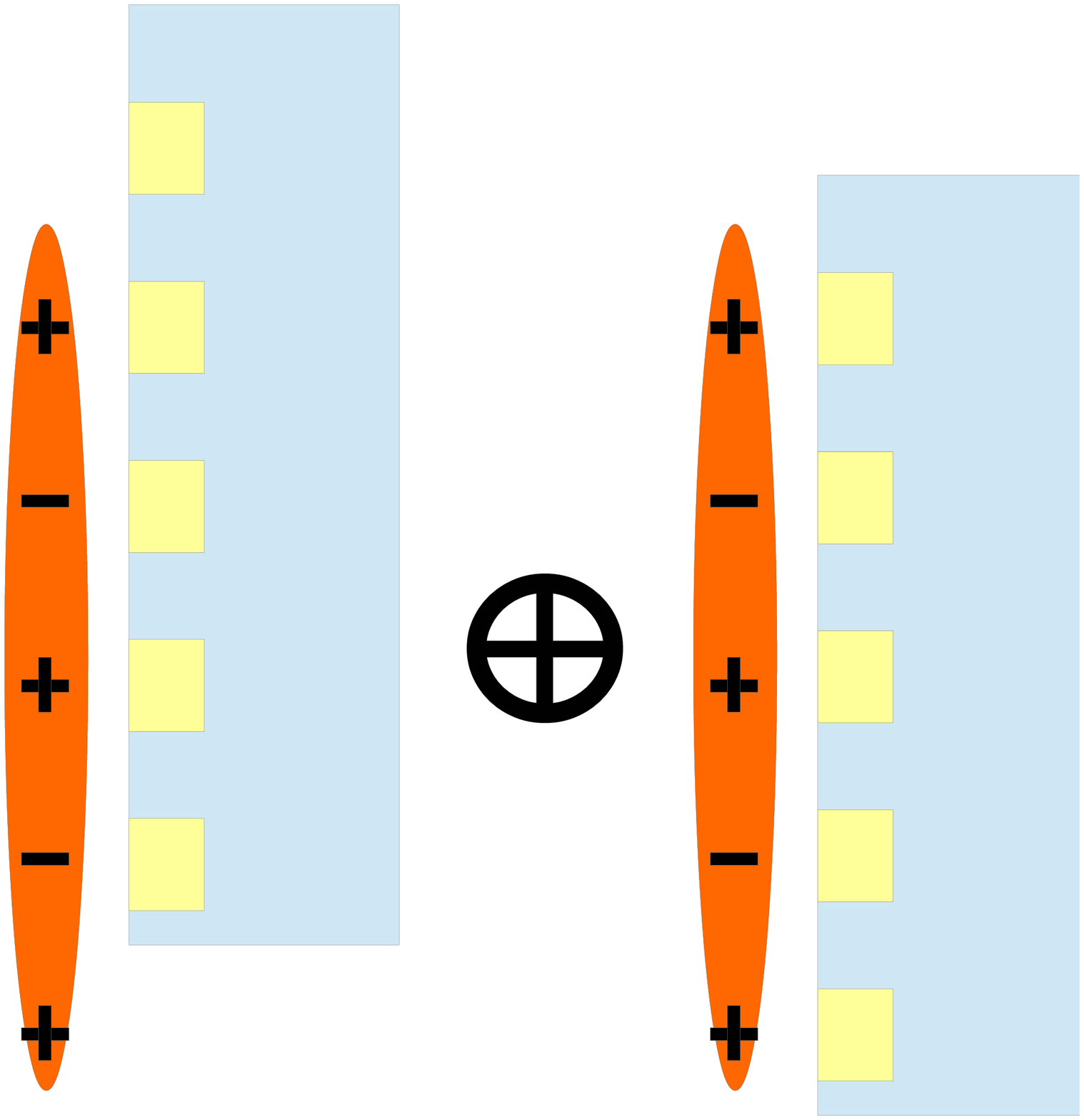}%
 \caption{\label{coupledmd}  A superimposed case of a measurement device with vertical delocalization and an incident wave packet.}
 \end{centering}
 \end{figure}

Let us now extend the above case of the heavy atoms to the case of a measurement device i.e. a screen, as in fig.\ \ref{broad}.  Here let us utilized a nearly monochromatic (wavelength $\lambda$) packet moving towards the screen but with a slow additional phase oscillation ($\lambda_{||}\ll\lambda$) parallel to the screen surface.  Let the screen have five adsorption sites and have separation equal to half this long wavelength oscillation $D=\lambda_{||}/2$ as in fig.\ \ref{verybroad}.  Now let the measurement device be delocalized in the vertical direction by a vertical shift $D$.  We consider this to be in the form of two narrow packets of equal amplitude akin to the case of the incident wave in fig.\ \ref{side}.  The resulting initial state is described by the sum of configurations in fig.\ \ref{coupledmd}.  
Upon interaction the sites on the screen now feel both a positive and negative amplitude component of the wave.  This is our first case of a correlated two body system.  The system slices into a set of 4+5=9 cases where the first four correspond to a screen that is upwardly displaced by $D$ and the other five do not.  For an ``observer'' living in the screen body itself, one of these cases appears to represent his initial data for the evolving future for all the initial data he has available to him.  If somehow these slices are brought together in his future and the photon fields radiated from the adsorption events are confined with the system, the neighboring slices can interfere and this would seem to be a statistical aberration that flows from an unknown source.  Now let us consider the situation from the ``meta-observer'' outside the system.  This person can interact with the screen before or after the screen interacts with the packet.  The bifurcation of amplitude gives the same results in both cases so there is nothing ``fuzzy'' about the measurement from the delocalized device and the measurement operations commute.

\section{Conclusions}
One alternate title to this article could have been: 
``The Cheap Photon and the Classical Limit: The Origin of Discrete State Machines, the Apparently 3D World, Quantum Measurement, the Arrow of Time and Why You Have Any Memory at All.''  It is impressive that such disparate topics should all be connected to mapping the classical world properly into quantum mechanics.  A sister document on the dynamic process of thermalization and time dependent fluctuations has also been recently completed by this author \cite{Chafin-auto}.  The many body wavefunction of a system is a complicated high dimensional object.  By including the photons a large number of degrees of freedom appear that allows condensing matter to sparsely occupy subdomains corresponding to very similar objects that retain independent existence for long periods of time.  This provides a subset of wavefunctions that correspond to classical bodies that can withstand many quantum slicing events without producing significant overlap.  The release of low mass particles from a condensed matter ``classical''  body leads to a product function state where the low mass component spreads rapidly and, when reabsorbed, creates a bifurcated class of such classical states with probabilities given by the Copenhagen interpretation defining a set of measurement events.  These are locations and times specified by the atomic granularity scale of our condensed matter and a temporal granularity scale by the photon decay process associated with binding times.  This resolves the paradoxes of quantum measurement and introduces an arrow of time in a rather simple fashion.  We have argued that the genesis of such a state follows naturally from early universe conditions assuming condensation of small clusters of very low internal energy have time to interact and produce the localized classicality that partition the wavefunction into Newtonian-like parts.  

One of the more unclear features yet to be resolved here is in the behavior of gases.  Gases are made of light particles that have rapid delocalization so the persistent localization property we have argued for solids is not applicable.  Collisions with solids surfaces of a container produce some localization by the slicing process but low diffusion rates suggest that this does not propagate well into the bulk of the gas.  Hydro and thermodynamic behavior either requires some regular interaction with condensed matter by collision or possibly by photons or by some other process.  We know that such gases have the power of producing quantum like measurement paths in cloud chambers (though clouds by definition involve condensed droplets).  These are not pointlike but line-like events.  This introduces an interesting direction to further investigate this model.  Ultracold gas dynamics has become a very popular probe of quantum limits on viscosity \cite{Dolfovo, Son}.  It is not clear that at such low temperatures for gases bound by fields and so not in contact with condensed matter, that hydro and thermo are valid limiting behaviors on any timescale.  These macroscopic formal models are often justified by vague scaling arguments.  It is hard to argue against them because we have lacked a proper quantum description of gases in its ``classical'' limit.  If this can be found, we may have a framework to see how well such a description can hold in the ultracold case and if such parameters like temperature and viscosity can have any relevant meaning for them.

\end{document}